\documentclass[superscriptaddress,twocolumn,nofootinbib,aps,prx]{revtex4-2} 
\usepackage[utf8]{inputenc}
\usepackage{comment}
\usepackage{graphicx} 
\usepackage[section]{placeins}
\usepackage{mathtools} 
\usepackage{physics} 
\usepackage{amsmath,dsfont,bm,physics} 
\usepackage{stmaryrd}
\usepackage{floatrow}
\usepackage[margin=1in]{geometry}
\usepackage[title]{appendix} 
\usepackage{blindtext} 
\usepackage{fancyhdr} 
\usepackage{indentfirst} 
\usepackage{natbib} 
\usepackage{xcolor}
\usepackage{txfonts}
\usepackage[export]{adjustbox}
\usepackage[colorlinks=true, linkcolor=blue, citecolor=blue, urlcolor=blue]{hyperref}
\usepackage[label font=bf,font=large,labelformat=simple,caption=false]{subfig}
\usepackage{ulem}

\begin{document}

\title{Generating Single- and Many-Body Quantum Magnonic States}
\author{V. Williams}
\email{willibte@bc.edu}
\affiliation{Department of Physics, Boston College, 140 Commonwealth Avenue, Chestnut Hill, Massachusetts 02467, USA}

\author{J. M. P. Nair}
\email{muttathi@bc.edu}
\affiliation{Department of Physics, Boston College, 140 Commonwealth Avenue, Chestnut Hill, Massachusetts 02467, USA}

\author{Y. Tserkovnyak}
\email{yaroslav@physics.ucla.edu}
\affiliation{Department of Physics and Astronomy and Bhaumik Institute for Theoretical Physics,
University of California, Los Angeles, California 90095, USA}

\author{B. Flebus}
\email{flebus@bc.edu}
\affiliation{Department of Physics, Boston College, 140 Commonwealth Avenue, Chestnut Hill, Massachusetts 02467, USA}

\date{\today}

\begin{abstract}
 The growing interest in quantum magnonics is driving the development of advanced techniques for generating, controlling, and detecting non-classical magnonic states. Here, we explore the potential of an ensemble of solid-state spin defects coupled to a shared magnetic bath as a source of such states. We establish a theoretical framework to characterize the quantum correlations among magnons emitted by the ensemble into the bath and investigate how these correlations depend on experimentally tunable parameters. Our findings show that the emitted magnons retain the quantum correlations inherent to the solid-state emitters, paving the way for the deterministic generation of quantum many-body magnonic states.
     
\end{abstract}

\maketitle

\textit{Introduction.}---Although still in its infancy, the field of quantum magnonics has recently gained momentum, driven by its  potential to advance quantum technologies~\cite{Yuan,Flebus1,Flebus2}. This growing interest is fueled by the unique and versatile properties of magnons, i.e., the quantized excitations of spin waves in magnetically ordered materials~\cite{rezende2020fundamentals}. Their broad tunability, intrinsic non-reciprocities, and seamless integration with established quantum platforms make them promising candidates for applications ranging from quantum information processing to scalable quantum communication networks. Central to the advancement of this field is the ability to generate, manipulate, and detect quantum magnon states, which would enable the encoding, transfer, and processing of information within hybrid quantum circuits~\cite{lachance2019hybrid,awschalom2021quantum} and serve as building blocks for continuous-variable quantum information processing \cite{Yuan,chumak2022advances}. Recent breakthroughs in encoding and transmitting information using magnons have inspired several theoretical proposals for generating single-magnon quantum states, most of which focus on enhancing the nonlinearities within magnonic systems such as magnon blockades~\cite{Yuan2,Haghshenasfard,Xie,Wang,Lachance-Quirion,Bittencourt,Xu}, and parametric pumping~\cite{Clausen,chumak2022advances}.

\begin{figure}[b]
\captionsetup{justification=justified,singlelinecheck=false}
\centering
\includegraphics[width=0.85\linewidth]{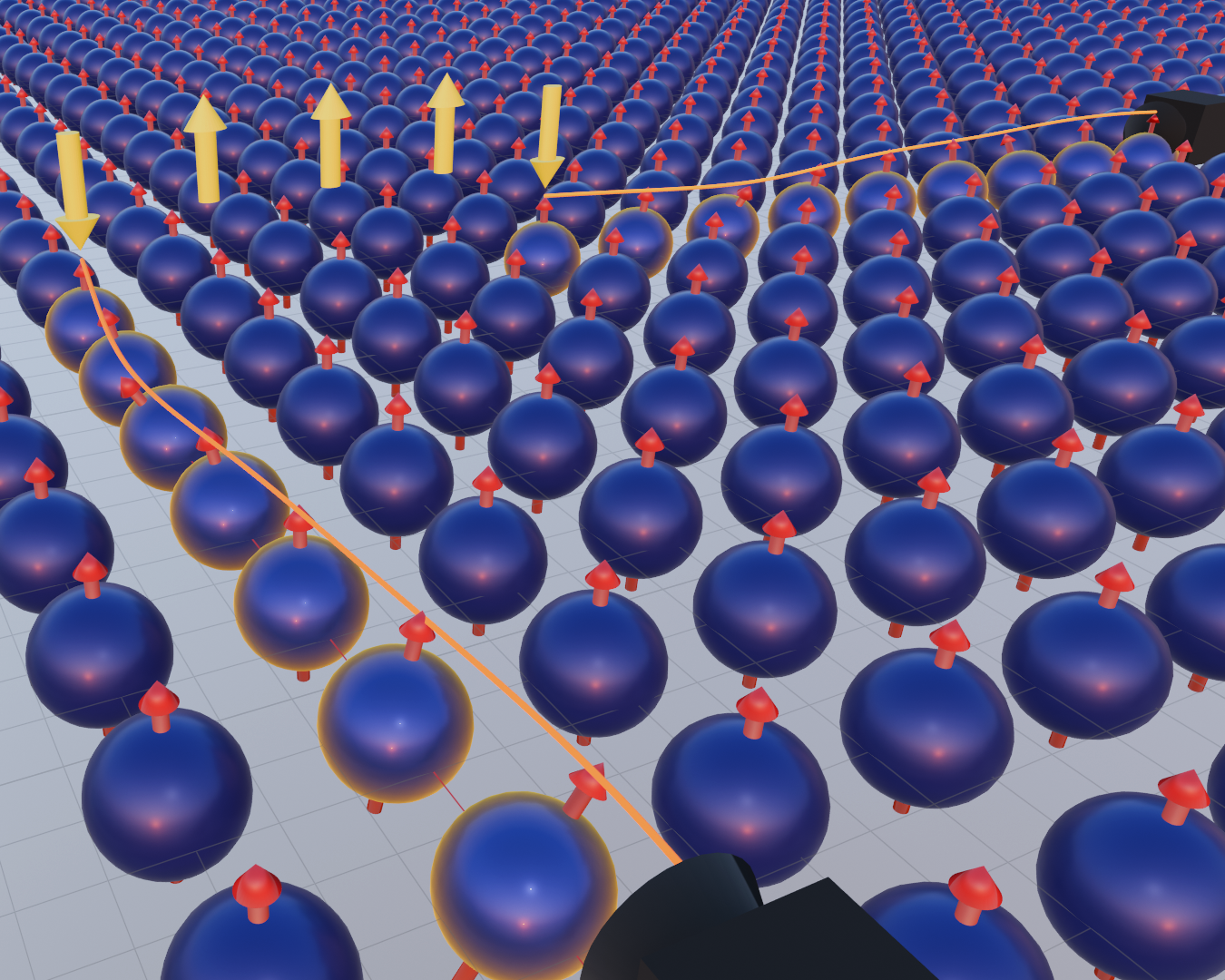}
    \caption{Generation of quantum-correlated magnonic states via spontaneous emission from an ensemble of solid-state defects coupled to a common ferromagnetic bath. The blue square lattice represents the magnet, with red arrows indicating ground-state spin polarization. Upon transitioning from the excited to the ground state, the 1D array of solid-state spin defects (yellow arrows) emits spin waves (orange line) into the magnetic bath, which is subsequently probed by the detectors (black camera icons).}   
    \label{pretty}
\end{figure}

In this Letter, we explore an alternative route for sourcing single magnon states that can also, more intriguingly, enable the generation of quantum many-body magnonic states. Our approach -- schematically illustrated in Fig.~\ref{pretty} -- relies on the well-established interaction between solid-state spin defects and the magnetic noise emitted by a nearby magnetic system. 
This qubit-bath coupling is routinely leveraged in quantum sensing experiments, which probe the properties of a magnetic system via their effect on the sensor spin’s relaxation time \cite{Du,HWang}. 
Quantum sensing experiments are typically conducted by setting the resonance frequency $\omega$ of the solid-state spin defect  above the magnetic system’s spin-wave gap $\Delta_\text{F}$ to maximize sensor-probe coupling strength. In this regime, i.e., $\hbar \omega > \Delta_\text{F}$, the solid-state spin defect can emit (absorb) magnons with energy $\hbar \omega$ into (or from) the magnetic bath, a process that can significantly shorten the spin defect's relaxation time. To further enhance the noise produced by thermal spin-wave fluctuations, the operating temperature $T$ is usually high enough to ensure 
a substantial thermal magnon population in the bath, i.e., $\Delta_\text{F} \ll k_B T$. 

The regime of interest to this work is instead the quantum one, i.e., $T \rightarrow 0$, corresponding to essentially no thermal magnons available for absorption by a spin defect.  In this scenario, if the spin defect is initialized in the excited state, its interaction with the magnetic system will eventually lead to the emission of a \textit{single magnon} -- a process that, to our knowledge, remains unexplored as a pathway for generating quantum magnonic states. Our goal is to build upon this potential of individual or uncorrelated solid-state spin defects as sources of single magnons to investigate the generation of quantum many-body magnonic states through ensembles of quantum-correlated solid-state spin defects.  A theoretical foundation for our exploration is provided by the recent work of Li and co-authors~\cite{Xin}, which has shown that a  magnetic bath, through its dynamical quantum fluctuations, can induce correlated dissipative dynamics in an ensemble of solid-state spin defects commonly coupled to it. These collective decay processes give rise to cooperative phenomena closely analogous to those established in light–matter platforms~\cite{Dicke1954,Chang,Facchinetti,Ana,Pivovarov,Reitz,Sipahigil,Walls,Plenio}, including the emergence of superradiant and subradiant dynamics.   In light--matter interfaces, however, the consequences of collective relaxation extend beyond the emitter dynamics themselves: correlations built within the emitter ensemble can be transferred to the outgoing radiation, where they become directly accessible through field-correlation measurements and can be harnessed for the generation of nonclassical propagating states~\cite{glauber1963quantum,Ana,wolf2020light}. It is therefore natural to ask whether the hybrid spin quantum platform depicted in Fig.~\ref{pretty} can enable an analogous form of correlation transfer. In this work, we address precisely this question by exploring whether correlations generated by the collective dynamics of the defect ensemble can likewise be encoded in the emitted magnon field, and if so, which observables reveal them. Our results show that many-body quantum correlations within the array can give rise to experimentally measurable, directionally correlated magnon emission, thereby opening a route toward the controlled generation and detection of nonclassical magnonic states.

\textit{Model.}---We begin by introducing the theoretical toolbox that describes the dynamics of an ensemble of solid-state spin defects interacting via the spin-wave fluctuations of a common ferromagnetic film.  For concreteness, we consider a 1D array oriented along the $\hat{\mathbf{x}}$ axis, with lattice constant $a_q$, of $N$ identical spins polarized along the $\hat{\mathbf{z}}$ axis, which captures the essential physics while remaining analytically tractable. Our framework is applicable to a wide range of solid-state spin defects. For instance, while  a nitrogen-vacancy (NV) center forms a spin-triplet system for low energies, at low temperatures and in the presence of a magnetic field $\mathbf{B}_{0}=B_0\mathbf{\hat{z}}$, it can be treated as an effective two-level system with frequency $\omega=\Delta_{0}-\tilde{\gamma} B_{0}$ in the $\{ \ket{0} , \ket{-1} \}$ subspace, where $\Delta_0$ is the NV center zero-field splitting and $\tilde{\gamma}>0$ (minus) its gyromagnetic ratio~\cite{Xin}.   The array is placed at a height $d$ above a ferromagnetic
film whose equilibrium spin density $\mathbf{s}(\boldsymbol{\rho})$  is aligned along the $\hat{\mathbf{z}}$ direction, i.e., $\mathbf{s} = ( s_x,  s_y, s)$, with $|\mathbf{s}| \simeq s$, with $s$ being the saturation spin density. We consider the experimentally relevant \textit{single-domain} regime stabilized by an external bias field \(B_0\), which—as demonstrated in hybrid NV–YIG platforms—can be reliably achieved in conventional films~\cite{Du,HWang}).
 The fluctuations of the homogeneous spin density around its equilibrium direction generate a magnetostatic field $\mathbf{B}_{\alpha} = \int d\boldsymbol{\rho}\;  \mathcal{D}(\boldsymbol{\rho}_\alpha,  \boldsymbol{\rho}, d) \mathbf{s}(\boldsymbol{\rho})$   where $\mathcal{D}$ is the tensorial magnetostatic Green’s function of the magnetic dipolar field \cite{MagGreen}.
 The Hamiltonian of the array can then be written  as
\begin{align}
\mathcal{H}=\mathcal{H}_\text{s}+\mathcal{H}_\text{int}\,, \;  \text{with} \; \; \mathcal{H}_\text{s} = -\frac{1}{2}\sum_\alpha \omega \sigma^z_\alpha\,, \nonumber \\
 \mathcal{H}_\text{int} = \tilde{\gamma}\sum_\alpha \left( \sqrt{2}B^+_\alpha \sigma^-_\alpha + \sqrt{2}B^-_\alpha \sigma^+_\alpha + B^z_\alpha \sigma^z_\alpha \right)\,,
\end{align}
where $B_\alpha^{\pm} = 
 B_\alpha^x \pm iB_\alpha^y$,  $\sigma^-_{\alpha} = \ket{0_\alpha}\bra{-1_\alpha}$ and  $\sigma^+_{\alpha} = \ket{-1_\alpha}\bra{0_\alpha}$, with $\boldsymbol{\sigma}_\alpha$ labeling the quantum spin  at the site $\boldsymbol{\rho}_{\alpha}$. 
After tracing out the magnonic degrees of freedom of the bath using a Born–Markov approximation \cite{Carmichael2,Walls,Reitz,Plenio}, one finds that, in the vanishing temperature limit, i.e., $T \rightarrow 0$, the evolution of the density matrix $\rho$ of the array can be written as \footnote{For a detailed derivation see \cite{Xin}.}  
\begin{equation}
    \begin{split}
        \frac{d \rho}{d t} = &-i\comm{\mathcal{H}_s + \underset{\alpha \ne \beta}{\sum}\left( J_{\alpha\beta}\sigma^+_\alpha \sigma^-_\beta \right)}{\rho} \\
        &+\underset{\alpha \beta}{\sum} \Gamma_{\alpha \beta} \left( \sigma^-_\beta \rho \sigma^+_\alpha - \frac{1}{2} \acomm{\sigma^+_\alpha \sigma^-_\beta}{\rho} \right)\,. \\
    \end{split}
    \label{master}
\end{equation}
Here, the term $\propto J_{\alpha \beta}$ represents a coherent exchange interaction that does not play a substantial role in our discussion as it is magnon number-conserving. The term $\propto \Gamma_{\alpha \beta}$, instead, describes dissipative processes, i.e., the emission of magnons   into  the ferromagnetic reservoir. For an array of non-interacting spins, the matrix $\boldsymbol{\Gamma}=(\Gamma_{\alpha \beta})$ becomes diagonal, with entry $\Gamma_{\alpha \alpha} = \Gamma_0$ corresponding to the relaxation rate of the isolated $\alpha$th spin. On the other hand, dissipative inter-qubit correlations result in finite off-diagonal terms $\Gamma_{\alpha \beta}$ that drive the correlated magnon emission and the ensuing cooperative quantum dynamics \cite{Xin}.

The explicit form of the matrix $\boldsymbol{\Gamma}$ depends on the spin susceptibility $\boldsymbol{\chi}$ of the magnetic film \cite{Xin,Ana,Casola,Masson},  which parameterizes its linear response to an external field  $\mathbf{h}$.  When $\hbar \omega > \Delta_F$, the main drive of correlated emission is the transverse magnetic noise encoded in the spin susceptibility $\chi^{+-}$ \cite{Xin,Flebus3}, which is defined through  
\begin{equation}
    s^{+}(\boldsymbol{\rho},t) = \gamma \int d\boldsymbol{\rho}'dt'\;  \chi^{+-}(\boldsymbol{\rho}-\boldsymbol{\rho}', t-t') h^+(\boldsymbol{\rho}', t')\,.
    \label{delta s}
\end{equation}
Here, $\gamma$ is the gyromagnetic ratio of the film, and $s^{\pm}=( s^{x}\pm i s^{y})/2$ and  $h^{\pm}=h^x \pm ih^y$.
In the long-wavelength limit, the transverse spin susceptibility can be extracted from the following Landau-Lifshitz-Gilbert (LLG) equation \cite{Landau} :
\begin{equation}
\frac{d \mathbf{s}}{d t} = -\gamma \mathbf{s} \cross \mathbf{B}_\text{eff} -\gamma  \mathbf{s} \cross \mathbf{h} + \frac{\alpha}{s}\mathbf{s} \cross \frac{d \mathbf{s}}{d t}\,,
\label{LLG}
\end{equation}
where $\alpha \ll 1$  is the dimensionless Gilbert damping parameter and $\mathbf{B}_\text{eff} = -\gamma^{-1}\partial \mathcal{H}_m/\partial \mathbf{s}$ is the effective (Landau-Lifshitz) field associated with the bath Hamiltonian. While Eq.~\eqref{master} holds for a generic stationary ferromagnetic bath with reciprocal spin-wave dispersion, i.e., $\omega_\text{F}(\mathbf{k})=\omega_\text{F}(-\mathbf{k})$  \cite{Xin24}, we focus here for concreteness on a
$U(1)$-symmetric ferromagnet~\cite{Xin}, whose Hamiltonian is given by 
 \begin{align}
    \mathcal{H}_m &= \!\!\int \!\!d\boldsymbol{\rho} \left[  -J\mathbf{s}(\boldsymbol{\rho}) \cdot \nabla^2 \mathbf{s}(\boldsymbol{\rho})  -    As^2_z(\boldsymbol{\rho}) + \gamma B_0s_z(\boldsymbol{\rho})   \right]\,\!, 
    \label{Hamiltonian}
    \end{align}
where $J$ and $A$ parametrize, respectively,  a nearest-neighbor Heisenberg exchange interaction and a uniaxial anisotropy. Equation~(\ref{Hamiltonian}) describes a ferromagnet with spin-wave dispersion  $\omega_\text{F}(k) = D k^2 + \Delta_\text{F}$, 
where $D = sJ$ is the spin stiffness and $\Delta_\text{F} = K + \gamma B_0$, with $K=2sA$,  the spin-wave gap. 
Plugging Eq.~\eqref{Hamiltonian} into Eq.~\eqref{LLG}, we obtain
\begin{equation}
    \chi^{+-}(k,\omega) = \frac{s/2}{D(k^2 - 1/\lambda^2) + i\alpha \omega}\,,
    \label{chi}
\end{equation}
where $\lambda = \sqrt{D/(\omega- \Delta_\text{F})}$ is the characteristic wavelength of the magnons emitted by the solid-state spin defects. Taking as an example an NV center ensemble interacting with a Yttrium Iron Garnet (YIG) film with spin stiffness $D=5.1\times10^{-28}\;\text{erg}\cdot\text{cm}^2$ and zero-field gap $K=3.6\times10^{-18}\;\text{erg}$, setting the external field to $B_0\sim 40$ mT,  results in a characteristic wavelength $\lambda 
\sim 280$ nm that  is much larger than the minimal NV-bath separation $d$
achievable experimentally~\cite{Xin}.  In this regime, i.e., $d \ll \lambda$,  the coherent and dissipative inter-qubit couplings can be written as
\begin{equation}
    J_{\alpha \beta}/\nu = -\frac{\pi}{2}\frac{\omega - \Delta_\text{F}}{\Delta_0} Y_0\left(\frac{\rho_{\alpha \beta}}{\lambda}\right)\,, \quad  \textcolor{blue}{\text{with}}  \, \, \alpha \neq \beta
  \label{8}  
\end{equation}
 and
\begin{equation}
    \Gamma_{\alpha \beta}/\nu = \pi\frac{\omega - \Delta_\text{F}}{\Delta_0} J_0 \left(\frac{\rho_{\alpha \beta}}{\lambda}\right) \textcolor{blue}{\Theta(\omega - \Delta_{\text{F}})}\,,
     \label{9}  
\end{equation}
with $\nu = \pi h^3(\gamma \tilde{\gamma})^2 s \Delta_0/D^2$ \cite{Xin}. Here, $\rho_{\alpha \beta}$ is the spatial separation between the $\alpha\:\!$th and $\beta\:\!$th qubits, and $J_0$ ($Y_0$) is the $0^\text{th}$ order Bessel function of the first (second) kind. 
By plugging Eqs.~\eqref{8} and~\eqref{9} into Eq.~(\ref{master}), one can calculate the relaxation dynamics of the NV-center ensemble for any given initial state. 
%

\textit{Correlated emission.}---In order to characterize the statistics of the magnons emitted by the qubit ensemble into the reservoir, one must first establish a metric for the detection of a single magnon. Drawing inspiration from the field of quantum optics~\cite{agarwal2012quantum}, we introduce the normalized second-order correlation function 
\begin{equation}
    g^{(2)}(\boldsymbol{\rho},\boldsymbol{\rho}'\!,t,t') =\! \frac{\expval{s^+(\boldsymbol{\rho},t) s^+(\boldsymbol{\rho}'\!,t') s^-(\boldsymbol{\rho},t) s^-(\boldsymbol{\rho}'\!,t')}}{\expval{s^+(\boldsymbol{\rho},t)  s^-(\boldsymbol{\rho},t)}\!\expval{ s^+(\boldsymbol{\rho}'\!,t') s^-(\boldsymbol{\rho}'\!,t')}}\, .
    \label{g22}
\end{equation}
 In the following, we focus on correlations in the far-field limit, which allows us to rewrite 
$g^{(2)}(\boldsymbol{\rho},\boldsymbol{\rho}'\!,t,t')$ as $g^{(2)}(\phi,\phi',t,t')$ where $\phi$ and $\phi'$ are the angles of the detectors measured from the
array axis. Furthermore, for the stationary fields considered in this work, we can further simplify  $g^{(2)}(\phi,\phi',t,t')\Rightarrow g^{(2)}(\phi,\phi',\tau,0)$, with $\tau= t -t'$. By invoking the Holstein-Primakoff transformation, i.e.,   $ s^+ \propto \sqrt{2s} a^{\dagger}$, where $a^{\dagger}$ is the magnon creation operator, Eq.~\eqref{g22} can be understood as the joint probability of two simultaneous magnon emissions normalized by their independent probabilities. In analogy with its quantum optical counterpart~\cite{glauber1963quantum,sudarshan1963equivalence}, it characterizes the nature of the quasi-particle source: coherent or thermal magnon sources are expected to exhibit $g^{(2)}(0)$ values of one or greater, while a value of $g^{(2)}(0)$ smaller than one is the hallmark of a single magnon source.

To compute Eq.~(\ref{g22}) explicitly, we reformulate it in terms of the dynamical evolution of the qubit ensemble described by Eq.~(\ref{master}). For $T \rightarrow 0$, the (circularly polarized) transverse fluctuations of the bath spin density~\eqref{delta s} can be defined as a response to the magnetic field generated by the qubit array, which reads as
\begin{align}
\mathbf{h}(\boldsymbol{\rho},t) = \sqrt{2} \tilde{\gamma} \overset{N}{\underset{\alpha=1}{\sum}} \mathcal{D}(\boldsymbol{\rho}_\alpha,\boldsymbol{\rho},d) \cdot\boldsymbol{\sigma}_{\alpha}(t)\,.
\label{NV field}
\end{align}
The real-space spin susceptibility $\chi^{+-}$ can be obtained by linearizing and Fourier transforming Eq.~(\ref{LLG}), resulting in the inhomogeneous Helmholtz equation :
\begin{align}
   \left[ \nabla^2 + k_0^2 - i\epsilon \right]   s^+(\boldsymbol{\rho},t) = -\frac{s\gamma}{D}\, h^+(\boldsymbol{\rho},t)\,,
   \label{helm}
\end{align}

\begin{figure*}[ht]
    \centering
    \sidesubfloat[]{\hspace{-20pt}\includegraphics[height=0.27\linewidth]{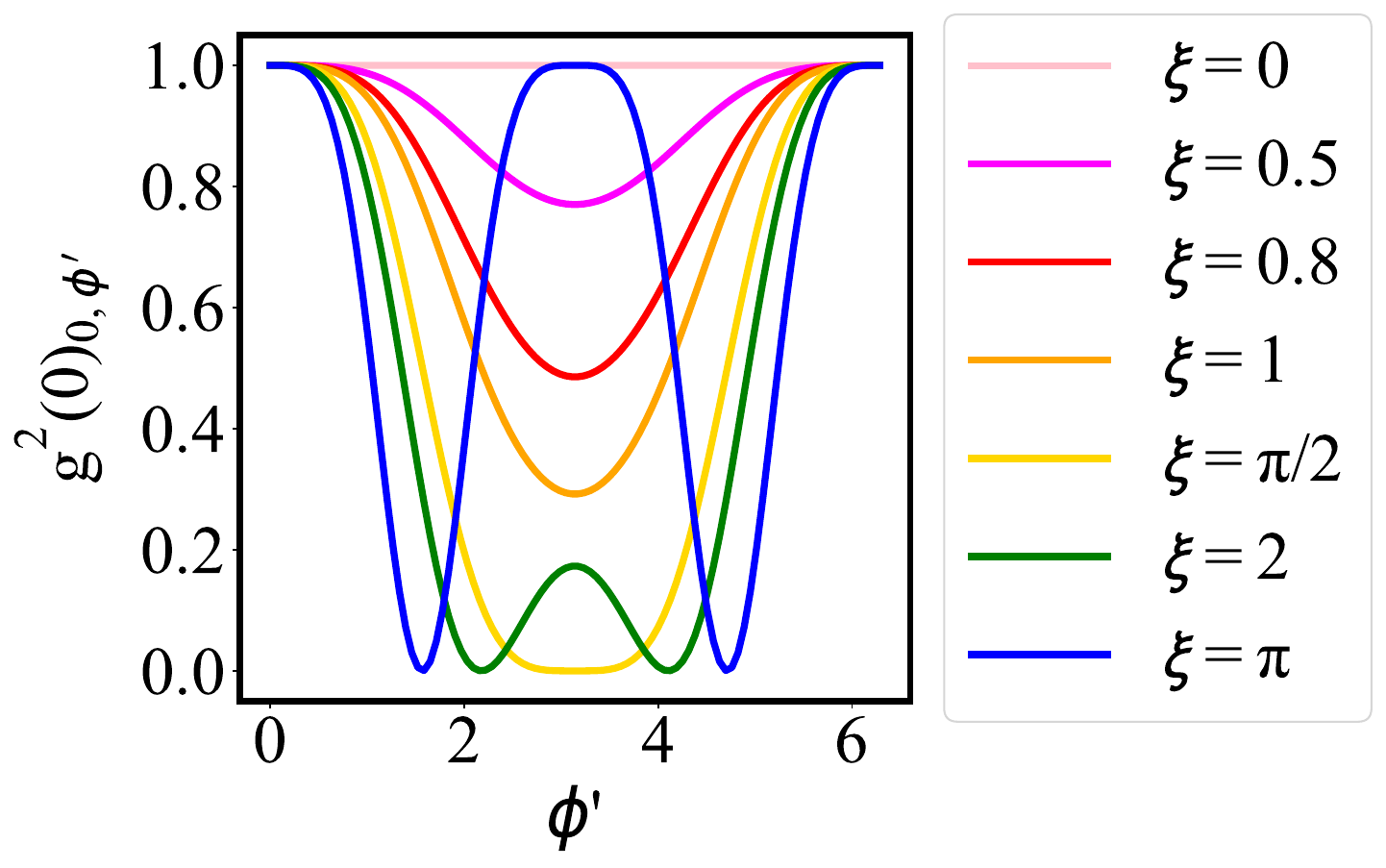}\hspace{4pt}}
    \sidesubfloat[]{\hspace{-10pt}\includegraphics[height=0.27\linewidth]{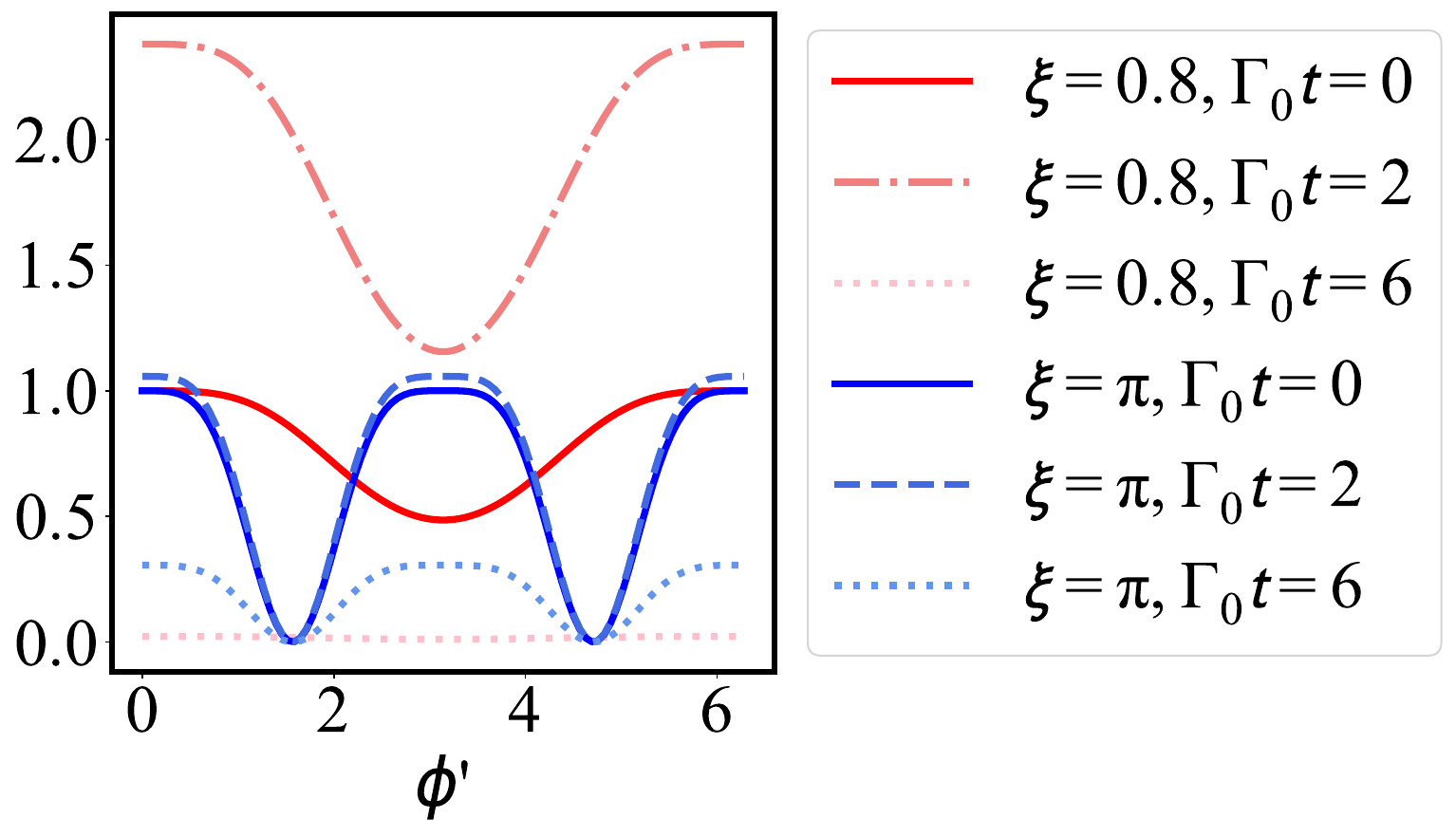}}
    \vspace{5pt}
    \sidesubfloat[]{\hspace{-25pt}\includegraphics[height=0.27\linewidth]{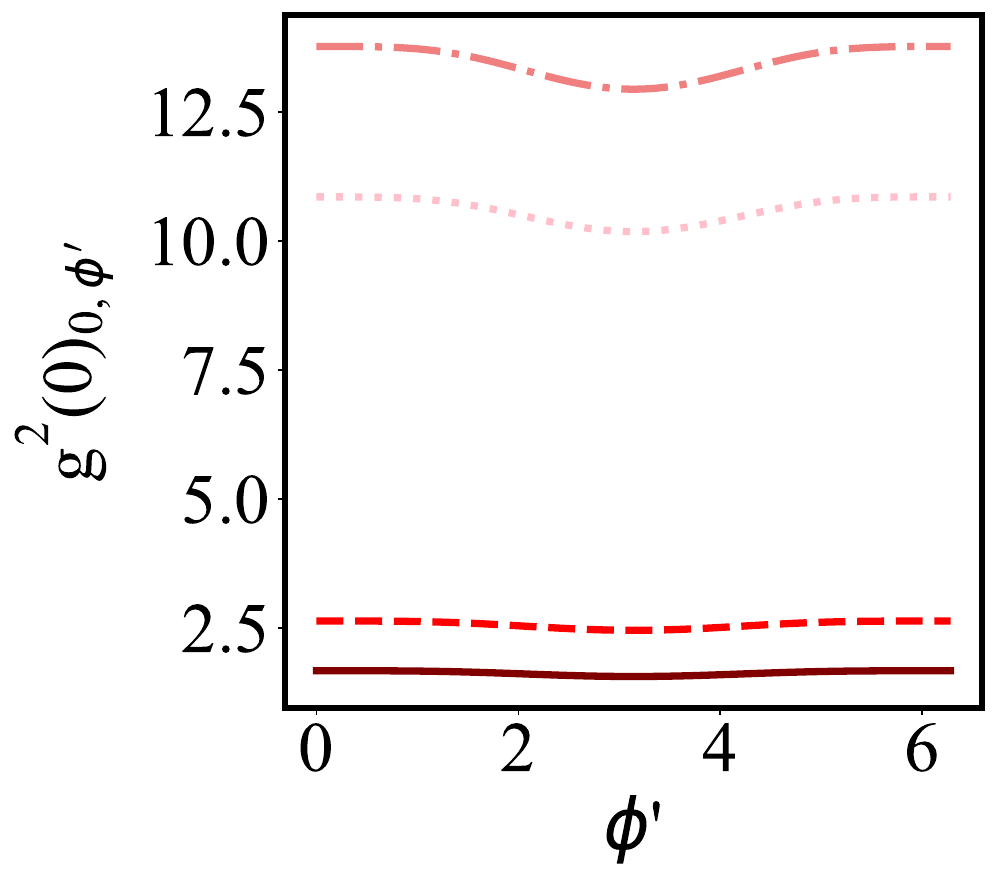}\hspace{4pt}}
    \sidesubfloat[]{\hspace{-10pt}\includegraphics[height=0.27\linewidth]{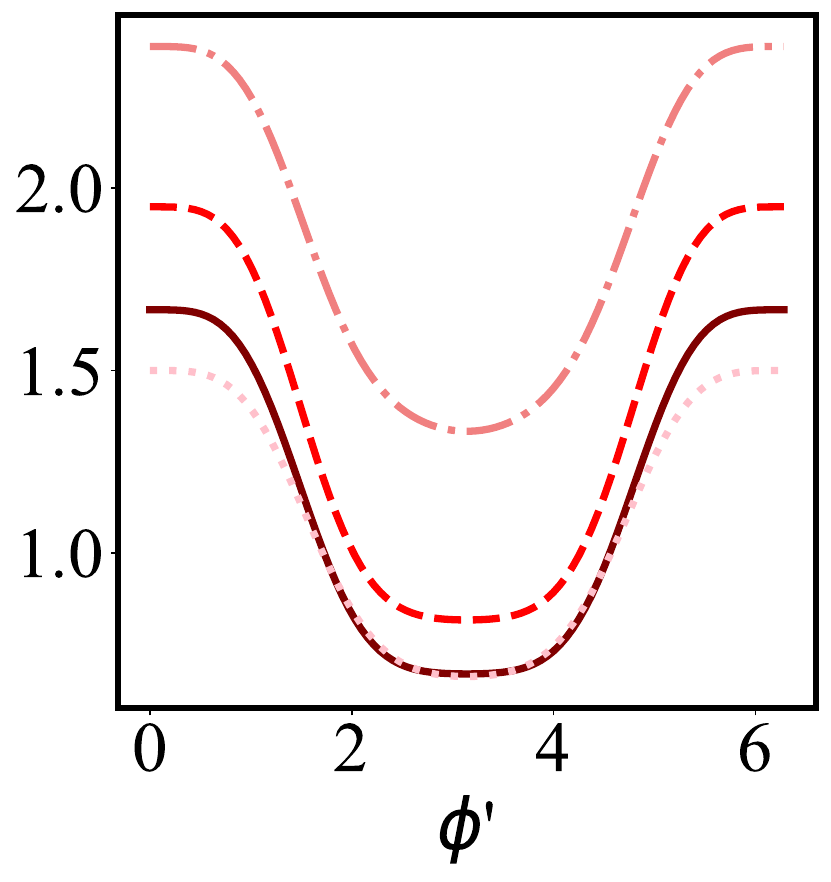}\hspace{4pt}}
    \sidesubfloat[]{\hspace{-2pt}\includegraphics[height=0.27\linewidth]{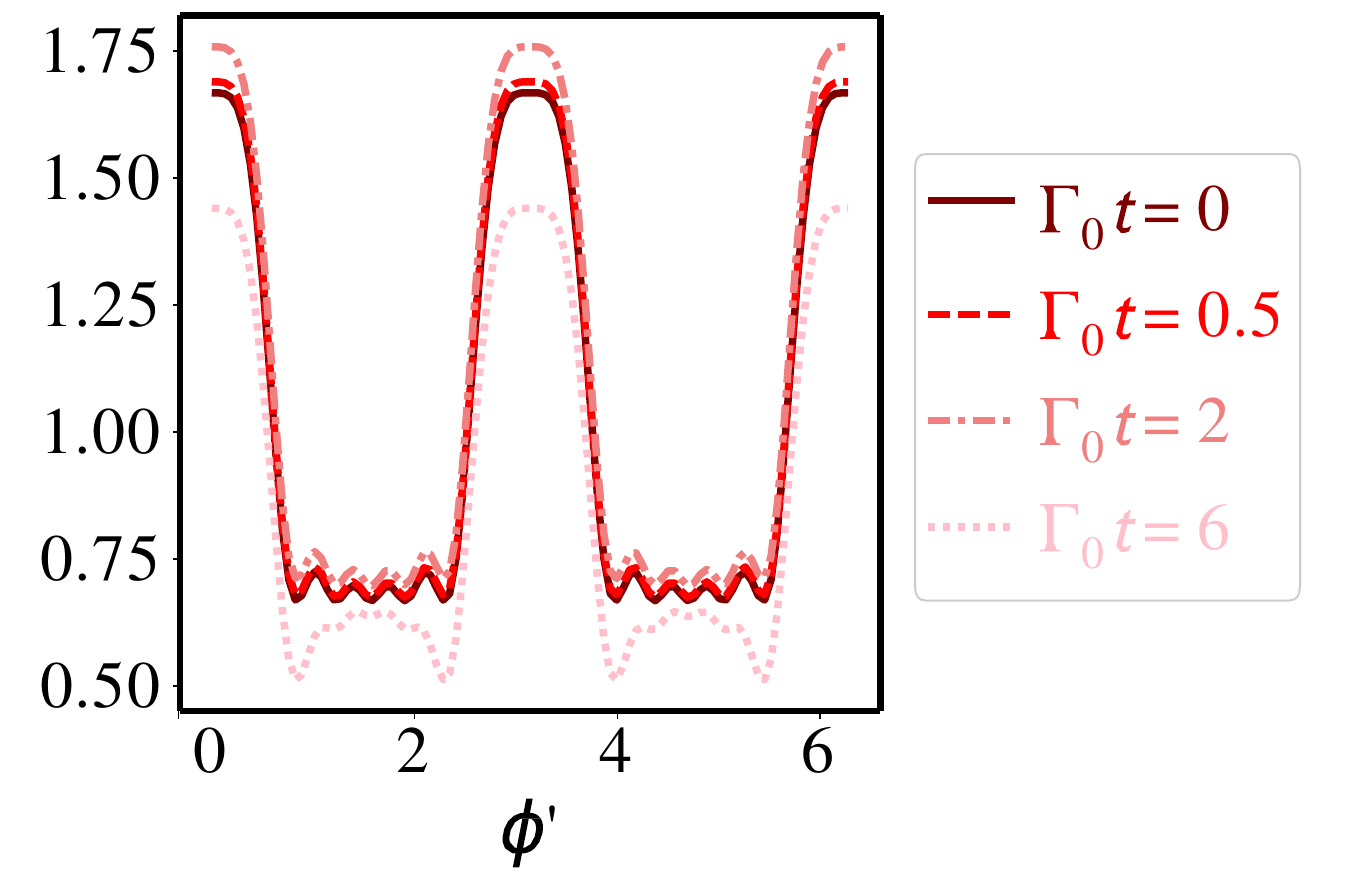}}
    \caption{Dependence of the zero-delay second-order correlation function~\eqref{g2final} on the emission angle $\phi'$, with fixed $\phi = 0$, for an ensemble of 2 emitters and different values of $\xi$: (a) at time $\Gamma_0 t \rightarrow 0$; (b) at various times. Dependence of the zero-delay second-order correlation function~\eqref{g22} on the emission angle $\phi'$, with fixed $\phi = 0$, for an ensemble of 6 emitters at times $\Gamma_0 t = 0$ (solid dark red), $\Gamma_0 t = 0.5$ (dashed red), $\Gamma_0 t = 2$ (dot-dashed light red), and $\Gamma_0 t = 6$ (dotted pink), for (c) $\xi = 0.1$; (d) $\xi = 0.5$; and (e) $\xi = \pi$. In all panels, the parameters correspond to an ensemble of NV-center spins interacting via a YIG thin film, as described in Ref.~\cite{Xin}.}
    \label{same}
   \FloatBarrier
\end{figure*}

\noindent with $k_0 = 1/\lambda$   and $\epsilon = \alpha \omega / D$.
When considering distances much less than the magnon decay length, Gilbert damping can be neglected, and  Eq.~(\ref{helm}) can be solved in the far field limit using a Green's function approach, yielding   
\begin{equation}
    \chi^{+-}(\boldsymbol{\rho},\boldsymbol{\rho}') = -i \sqrt{\frac{1}{8\pi k_0 \rho}}e^{ ik_0\rho}e^{-i\frac{\pi}{4}} e^{-ik_0\hat{\boldsymbol{\rho}}\cdot \boldsymbol{\rho}'}\,,
    \label{Green}
\end{equation}
where $\chi^{+-}(\boldsymbol{\rho},\boldsymbol{\rho}')$ denotes the outgoing plane-wave solution to Eq. (\ref{helm}).
Using Eqs.~(\ref{delta s}),~(\ref{NV field}), and~(\ref{Green}), we can rewrite
\begin{equation}
    s^+(\boldsymbol{\rho},t) =  i\frac{\gamma \tilde{\gamma}s}{D} \sqrt{\frac{k_0}{8\pi\rho}}e^{-k_0d} \overset{N}{\underset{\alpha=1}{\sum}} e^{-i k_0 \Hat{\boldsymbol{\rho}} \cdot \boldsymbol{\rho}_\alpha}       \sigma^+_\alpha(t),
    \label{fluc}
\end{equation}
where
$\cos^{-1}{(\hat{\boldsymbol{\rho}}\cdot \hat{\boldsymbol{\rho}}_\alpha)} = \phi$ is the angular position of the detector in the plane of the ferromagnetic film.
By plugging Eq.~\eqref{fluc} into Eq.~\eqref{g22}, the 
second-order correlation function can be evaluated by numerically solving Eq.~\eqref{master}~\cite{QuTip}.

\textit{Results.}---While our framework applies to an arbitrary number $N$ of magnon emitters, we find that a small number of solid-state spin defects is sufficient to illustrate the salient features of the correlated emission. For $N=2$, plugging Eq.~\eqref{fluc} into Eq.~(\ref{g22}) and setting $\tau=0$ yields the zero-delay correlation function  
\begin{equation}
\begin{split}
        g^{(2)}(0)_{\phi,\phi'} &= \frac{(1 + q(\xi,\phi,\phi') )\expval{\sigma^+_1 \sigma^+_2 \sigma^-_1 \sigma^-_2}}{2\left|\expval{\sigma^+_1\sigma^-_1}\right|^2}, \\
        \end{split}
        \label{g2final}
\end{equation}
where \( q(\xi, \phi, \phi') = \cos(\xi \cos\phi - \xi \cos\phi') \) and \( \xi = k_0 a_q \) with inter-qubit spacing $a_q$. The correlators \( \expval{\sigma^+_1 \sigma^+_2 \sigma^-_1 \sigma^-_2} \) and \( \expval{\sigma^+_1 \sigma^-_1} \) quantify, respectively, the joint transition probability associated with two correlated magnon emissions - one originating from NV\(_1\) and the other from NV\(_2\) - and the probability of two independent magnon emissions.


 
Figure~\ref{same}(a) displays the angular dependence of the zero-delay second-order correlation function~\eqref{g2final} as a function of the normalized interqubit separation \( \xi \), evaluated in the early-time limit \( \Gamma_0 t \rightarrow 0 \), where both correlators \( \langle \sigma^+_1 \sigma^+_2 \sigma^-_1 \sigma^-_2 \rangle \) and \( \langle \sigma^+_1 \sigma^-_1 \rangle \) approach unity. Although no magnons have yet been emitted, \( g^{(2)}(0)_{\phi,\phi'} \) captures the angular structure of the joint emission amplitude, governed by interference between spin waves of wavelength \( \lambda \). In the Dicke limit (\( \xi \to 0 \)), where the emitters are nearly indistinguishable, the emission becomes fully isotropic, resulting in \( g^{(2)}(0) = 1 \), independent of direction (red line). As \( \xi \) increases, spatial phase differences accumulate between emitters, giving rise to directional interference effects. In particular, when \( \xi(\cos\phi - \cos\phi') = \pi \), destructive interference fully suppresses the joint emission probability, yielding \( g^{(2)}(0)_{\phi, \phi'} = 0 \).


To investigate the role of the quantum-correlated dynamics of the ensemble, we explore the dynamical evolution of the magnonic correlations~\eqref{g2final} for a strongly (weakly) correlated, i.e., $\xi = 0.8$ ($\xi = \pi$) ensemble. As shown in Fig.~\ref{same}(b), the strongly correlated ensemble exhibits pronounced bunching at early times, followed by a transition to strong anti-bunching at later times. This behavior reflects cooperative dynamics within the array: the initial bunching indicates an enhanced likelihood of simultaneous magnon emission due to collective decay processes, while the subsequent anti-bunching arises from excitation depletion and temporal correlations. 
In contrast, the correlation function for the weakly correlated ensemble remains nearly unchanged during the evolution, closely resembling the angular structure shown in Fig.~\ref{same}(a), which indicates that interference predominantly governs its angular dependence.

Finally, we simulate Eq.~(\ref{master}) for an ensemble of \( N = 6 \) emitters to examine whether the qualitative features identified in the \( N = 2 \) ensemble persist in larger systems. As in the two-qubit case, signatures of quantum-correlated dynamics emerge at small inter-emitter spacing. For \( \xi = 0.1 \), Fig.~\ref{same}(c) reveals pronounced bunching at early times, indicative of cooperative behavior driven by collective decay. This effect weakens with increasing separation, as shown in Fig.~\ref{same}(d) for \( \xi = 0.5 \). In contrast, for \( \xi = \pi \), Fig.~\ref{same}(e) shows that the correlations remain essentially static in time, and the angular structure is dominated by interference, with no evidence of dynamical enhancement in the emission profile.

\textit{Discussion and conclusions.}---In this work, we identify a regime in which solid-state spin defects coupled to a common magnetic system can serve as a source of both single- and many-body quantum magnonic states. We develop a general framework to characterize the statistics of magnons emitted by the correlated qubit ensemble, which enables us to explore how the signatures of correlated qubit decay are imprinted onto the emitted magnons. Specifically, focusing on a two-qubit setup, we show how the interference between the emitted spin waves and the quantum correlations of their sources can be probed in the bath via detection measurement. The detection of correlated magnon emission could be explored through hybrid qubit--magnon architectures such as the single-magnon counting scheme demonstrated in Ref.~\cite{Nakamura20}, which provides a realistic route toward time- and angle-resolved measurements of $g^{(2)}(\varphi,\varphi')$. While our analysis focuses on the interactions between solid-state defects and a simple ferromagnet at thermal equilibrium, our framework can be straightforwardly generalized to baths engendering more complex quantum dynamics~\cite{nair2024reservoirengineeredspinsqueezingquantum,Xin24}. For instance, beyond reciprocal baths, coupling the emitter array to surface Damon--Eshbach modes could enable the generation of chiral, direction-selective many-body magnonic states, while domain walls in magnetic films---acting as one-dimensional magnonic waveguides---may provide a particularly promising setting for observing and manipulating such collective dynamics, as their guided modes confine and steer correlated excitations along narrow channels \cite{flebus2018,Finco2021}, thereby enhancing their detectability and coherence.
A natural extension of our framework is to consider defect arrays in higher-dimensional geometries (e.g., 2D). In such cases, the interference pattern of the emitted magnons would reflect the new geometry, while the correlated-emission mechanism itself remains essentially the same. A systematic study of correlated emissions in 2D arrays lies beyond the scope of the present work and will be pursued in future investigations.

Overall, our proposal opens a novel pathway for generating quantum magnonic states and emphasizes the need for future investigations to establish a clear link between the quantum correlations in the collective dynamics of the qubit ensemble and a measurable quantification of entanglement in the emitted magnons.

\textit{Acknowledgments.}---The authors thank X. Li, J. Marino, and D. Chang for helpful discussions, and E. Foisy for the production of illustrations. B.F. acknowledges support from the ONR under Grant No. N000142412427 and Y.T. acknowledges support from the NSF under Grant No. DMR-2049979.




\bibliography{sample}

\end{document}


\title{Supplemental Material for: \\[2pt]
\emph{Generating single- and many-body quantum magnonic states}}
\author{V. Williams}
\email{willibte@bc.edu}
\affiliation{Department of Physics, Boston College, 140 Commonwealth Avenue, Chestnut Hill, Massachusetts 02467, USA}

\author{J. M. P. Nair}
\email{muttathi@bc.edu}
\affiliation{Department of Physics, Boston College, 140 Commonwealth Avenue, Chestnut Hill, Massachusetts 02467, USA}

\author{Y. Tserkovnyak}
\email{yaroslav@physics.ucla.edu}
\affiliation{Department of Physics and Astronomy and Bhaumik Institute for Theoretical Physics,
University of California, Los Angeles, California 90095, USA}

\author{B. Flebus}
\email{flebus@bc.edu}
\affiliation{Department of Physics, Boston College, 140 Commonwealth Avenue, Chestnut Hill, Massachusetts 02467, USA}

\date{\today}

\maketitle

This Supplemental Material is organized as follows. 
Section~S1 starts from the linearized and fourier-transformed spin susceptibility defined in Eq.~(\textcolor{blue}{4}) of the main text and derives the magnon field operator in Eq.~(\textcolor{blue}{13}).
Section~S2 discusses the time dependence of Eq.~(\textcolor{blue}{9}) and its consistency with approximations (far field) made in the derivation of Eq.~(\textcolor{blue}{12})-(\textcolor{blue}{13}).

\section*{S1.  Derivation of Magnon Field Operator}
\setcounter{equation}{0}

We begin by defining the LLG equation as in Eq. (\textcolor{blue}{4}),

\begin{equation}
\frac{d \mathbf{s}}{d t} = -\gamma \mathbf{s} \cross \mathbf{B}_\text{eff} -\gamma  \mathbf{s} \cross \mathbf{h} + \frac{\alpha}{s}\mathbf{s} \cross \frac{d \mathbf{s}}{d t}\,,
\label{LLG}
\end{equation}
where $\mathbf{s}= \delta s^x \Hat{x} + \delta s^y \Hat{y} + s \Hat{z}$ is the spin density of the ferromagnetic film, expanded around its equilibrium axis, $\hat{z}$ and $\mathbf{B}_\text{eff}$ is the effective magnetic field generated by the Hamiltonian of Eq. (\textcolor{blue}{5}). Fourier transforming Eq. \eqref{LLG} temporally and making use of circular basis, to write the transverse fluctuations as $s^{\pm}=(\delta s^{x}\pm i \delta s^{y})/2$  and the applied field as $h^{\pm}=h^x \pm ih^y$, yields
\begin{equation}
     (1-i\alpha)\omega_+ -(s J\nabla^2 - 2As - \gamma B_0 ) s^+ =  \gamma sh^+,
     \label{LLG2}
\end{equation}
for the right-circular mode. Here, $\omega_+$ corresponds to the positive frequency of the right handed mode and relates to the left handed mode as $\omega_+ = - \omega_-$. This corresponds to the usual circular basis definition after the Fourier transform, redefining the conjugate as $\chi^{+-}(\omega_+) =\chi^{-+}(\omega_-)^*$, which is the typical convention.
Rearranging Eq. \eqref{LLG2} in a more standard notation gives
\begin{equation}
   \left[ \nabla^2 - \frac{(1-i\alpha)\omega_+}{sJ} - \frac{2A}{J} - \frac{\gamma B_0}{sJ} \right]  s^+ =  -\frac{\gamma s}{sJ} h^+,
   \label{Helm1}
\end{equation}
which corresponds with $\chi^{+-}(\omega_+)$. We then define both $\chi^{+-}$ and $\chi^{-+}$ in terms of $\omega_-$ as it describes the emissions, i.e. $\omega>\Delta_F$, with the real characteristic magnon wavelength $\lambda (\omega_-)$ rather than the non-resonant and non-propagating $\lambda(\omega_+) = \lambda'$. Substituting $\omega_-$ into Eq. \eqref{Helm1} yields
\begin{equation}
   \left[ \nabla^2 + \frac{(1-i\alpha)\omega_-}{sJ} - \frac{2A}{J} - \frac{\gamma B_0}{sJ} \right]  s^+ =  -\frac{\gamma s}{sJ} h^+.
   \label{Helm2}
\end{equation}
Defining the spin stiffness $D=sJ$, anisotropy field $K=2sA$, spin wave gap $\Delta_\text{F} = K+\gamma B_0$, and redefining $\omega_- \rightarrow \omega$ and plugging them into Eq. \eqref{Helm2} simplifies to the 2D Helmholtz equation
\begin{align}
   \left[ \nabla^2 + k_0^2 - i\epsilon \right]   s^+(\boldsymbol{\rho},t) =  -\frac{s\gamma}{D}\, h^+(\boldsymbol{\rho},t)\,,
   \label{helm}
\end{align}
where $\boldsymbol{\rho}$ is a poisition within the film, $t$ is time, $\epsilon = \alpha \omega / D$, and $k_0 = \sqrt{(\omega - \Delta_\text{F})/D}$, corresponding to the magnon wavelength as $k_0=1/\lambda$. Equation \eqref{helm} can be solved using the Green's function 
\begin{equation}
    G(\boldsymbol{\rho},\boldsymbol{\rho}') = -\frac{i}{4} H_0^{(1)} (k_0|\boldsymbol{\rho}-\boldsymbol{\rho}'|),
    \label{green1}
\end{equation}
where
\begin{equation}
    H_0^{(1)}(z) \approx \sqrt{\frac{2}{\pi z}}e^{i(z-\frac{\pi}{4})}, z\xrightarrow{} \infty,
    \label{hank}
\end{equation}
is the zeroth order Hankel function of the first kind which corresponds to an outwardly propagating wave with azimuthal invariance. In the main text, the response function is denoted by \( \chi(k, \omega) \), whereas here we instead write the real-space Green's function \( G(\boldsymbol{\rho},\boldsymbol{\rho}')\) and its momentum-space form \( G(k, \omega) \), evaluated at fixed frequency $\omega$, to emphasize their role as the Green's function kernels of Eq. \eqref{helm}. Plugging Eq. \eqref{hank} into Eq. \eqref{green1} gives the Green's function
\begin{equation}
    G(\boldsymbol{\rho},\boldsymbol{\rho}') = -i \sqrt{\frac{1}{8\pi k_0|\boldsymbol{\rho}-\boldsymbol{\rho}'|}}e^{ik_0|\boldsymbol{\rho}-\boldsymbol{\rho}'|}e^{-i\frac{\pi}{4}}.
    \label{green2}
\end{equation}
This can be further simplified by considering the far-field approximation ($\rho'\ll \rho$)

\begin{equation}
    |\boldsymbol{\rho}-\boldsymbol{\rho}'|/\rho \approx 1-\Hat{\boldsymbol{\rho}}\cdot \boldsymbol{\rho}'/\rho + \order{\left(\rho'/\rho \right) ^2 }.
    \label{farfield}
\end{equation}
Using Eq. \eqref{farfield}, the Green's function can be written

\begin{equation}
    G(\boldsymbol{\rho},\boldsymbol{\rho}') = -i \sqrt{\frac{1}{8\pi k_0 \rho}}e^{ik_0\rho}e^{-i\frac{\pi}{4}} e^{-i\mathbf{k}'\cdot \boldsymbol{\rho}'}
    \label{greenFinal}
\end{equation}
where $\mathbf{k}' = k_0 \Hat{\boldsymbol{\rho}}$. Eq. \eqref{greenFinal} corresponds directly with Eq. (\textcolor{blue}{12}) in the main text. Finally, we can use Eq. \eqref{greenFinal} to write the solution to the 2D Helmholtz equation, i.e. the response 
\begin{equation}
    s^+(\boldsymbol{\rho},t) =  \gamma \int d\rho' G(\boldsymbol{\rho},\boldsymbol{\rho}') h^+(\boldsymbol{\rho}',t),
    \label{s}
\end{equation}
where the time dependence of $s^+(\boldsymbol{\rho},t)$ is identical to that of the source field $h^+(\boldsymbol{\rho}',t)$. Finally, plugging Eqs. \eqref{greenFinal} and (\textcolor{blue}{10}) into Eq. \eqref{s} yields Eq. (\textcolor{blue}{13}).

\section*{S2. Discussion on Time Parameters}
To avoid ambiguity, we briefly clarify the time dependence of the second-order coherence function. In particular, we separate the contributions coming from operator evolution and those arising from the density matrix dynamics. The second-order correlation function is defined in Eq. (\textcolor{blue}{9}) of the main text as 
\begin{equation}
    g^{(2)}(\boldsymbol{\rho},\boldsymbol{\rho}'\!,t,t') =\! \frac{\expval{s^+(\boldsymbol{\rho},t) s^+(\boldsymbol{\rho}'\!,t') s^-(\boldsymbol{\rho},t) s^-(\boldsymbol{\rho}'\!,t')}}{\expval{s^+(\boldsymbol{\rho},t)  s^-(\boldsymbol{\rho},t)}\!\expval{ s^+(\boldsymbol{\rho}'\!,t') s^-(\boldsymbol{\rho}'\!,t')}}\, ,
    \label{g22}
\end{equation}
where $t $ and $ t'$ are the operator times for two subsequent magnon emissions. In general, the operator time dependence goes as $\sigma^\pm(t) = \sigma^\pm e^{i\omega_\pm t}$, such that any observable combination of operators will be stationary so long as $\omega_+ = - \omega_-$, as is the case here.  This allows for a simplification to $\tau = t-t'$ which just represents the delay between the emission (or detection) of one magnon and the other. Stationarity makes the operator time evolution insensitive to whether the time argument refers to emission or detection; the retardation shift from the far-field limit similarly does not modify the correlator, making only their separation $\tau$ physical. In this sense, the only remaining time parameter in Eq. \eqref{g22} is the one implicit in the state $\expval{\cdots}$ being passed on by the evolution of the density matrix. When we refer to ‘early times’ in the main text, we mean early in the evolution of the array, i.e., the density matrix; this allows for the far-field regime at the operator level to coexist with the early-time regime at the density matrix level.